# Relative phase stability and lattice dynamics of $NaNbO_3$ from first-principles calculations

R.Machado, M. Sepliarsky and M.G. Stachiotti

*Instituto de Física Rosario, Universidad Nacional de Rosario, 27 de Febrero 210 Bis, (2000) Rosario, Argentina.*

*Abstract*

We report total energy calculations for different crystal structures of $NaNbO_3$ over a range of unit cell volumes using the all-electron full-potential (L)APW method. We employed both the local-density approximation (LDA) and the Wu-Cohen form of the generalized gradient approximation (GGA-WC) to test the accuracy of these functionals for the description of the complex structural behavior of $NaNbO_3$. We found that LDA not only underestimates the equilibrium volume of the system but also predicts an incorrect ground state for this oxide. The GGA-WC functional, on the other hand, significantly improves the equilibrium volume and provides relative phase stability in better agreement with experiments. We then use the GGA-WC functional for the calculation of the phonon dispersion curves of cubic $NaNbO_3$ to identify the presence of structural instabilities in the whole Brillouin zone. Finally, we report comparative calculations of structural instabilities as a function of volume in $NaNbO_3$ and $KNbO_3$ to provide insights for the understanding of the structural behavior of $K_{1-x}Na_xNbO_3$ solid solutions.

## 1. Introduction

$NaNbO_3$ (NN) is probably the most complex cubic perovskite known. It has an antiferroelectric (AFE) structure at room temperature and exhibits an unusual complex sequence of temperature driven structural phase transitions which are not clearly understood [1]. The high-temperature phase is the simple prototype cubic structure as in the other $ABO_3$ perovskites. Below 913K, a whole series of structural phase transitions has been found and at least six more phases have been identified. The successive phases, which are labelled $T_2$ , $T_1$ , S, R and P, are of orthorhombic symmetry except $T_2$ which is tetragonal. Between the four orthorhombic phases present in the temperature range 848–173K, the R phase is the most complicated one, having a unit cell containing 24 formula units. All these phases involve octahedral rotations or tilts. From 638K down to at least 170K, $NaNbO_3$ is antiferroelectric with an orthorhombic unit cell containing eight formula units. At low temperature a ferroelectric (FE) phase N with rhombohedral symmetry occurs below 170K.

The complexity of the structural phase-transition sequence of $NaNbO_3$ suggests the presence of several competing structural instabilities with very similar free energies. For that reason, the phase transition sequence is still not certain, and the symmetry of several phases remains controversial. For example, using high-resolution neutron scattering, Darlington and Knight [2] have shown that the symmetry of the room temperature phase P should be monoclinic rather than orthorhombic. An additional phase transition at 463K was proposed [3], whereas new phases can be induced by doping or applying an electric field [4–6]. Moreover, it was shown that the temperature characteristics of transitions has a crystalline-size dependence [7] ; and a recent study indicates that the FE rhombohedral N phase might coexist with the AFE orthorhombic P phase at low temperatures [1] .

The structural complexity of $NaNbO_3$ has implications when the material is alloyed with other pervoskites, as in the technologically important $K_{1-x}Na_xNbO_3$ solid solutions. Among various candidates for lead-free piezoelectric materials, ceramics based on potassium sodium niobate (abbreviated as KNN) are the most promising, because of its outstanding piezoelectric and ferroelectric properties as well as its better environmental compatibility than other lead-free piezoelectric ceramics. The x=0.5 composition modified by Li and Ta [8] has proven to be a piezoelectric with properties almost comparable to their lead-based counterpart, $PbTi_xZr_{1-x}O_3$ (PZT). $KNbO_3$ exhibits the same sequence of phase transitions as $BaTiO_3$ , transforming from the cubic paraelectric to the tetragonal phase at 708K, from the tetragonal to the orthorhombic phase at 498K and from the orthorhombic to the rhombohedral phase at 263K. The tetragonal, orthorhombic and rhombohedral phases are all ferroelectric. The doping of $KNbO_3$ with Na leads to a system with many phase boundaries. Over most of the phase diagram the phase transition sequence is the same as in pure $KNbO_3$ and, except for the orthorhombic-rhombohedral, the phase transition temperatures vary little over the whole composition range (Curie temperature around 700K). There is a morphotropic phase boundary near the x=0.5 composition separating two orthorhombic phases, so this composition is therefore considered as a viable alternative to those materials for selected applications [9]. However, the phase diagram of KNN has been recently updated showing that this system does not display a morphotropic phase boundary comparable to the one observed in PZT, and the most significant structural change as a function of composition occurs because of the change of the tilt system [10].

Given this experimental background, first-principles studies of $NaNbO_3$ are highly desirable, since this computational approach is an extremely valuable tool to identify competing structural instabilities in perovskites. In spite of the importance of phonons in the physics of ferroelectrics, there has been no study of the phonon dispersion relation of $NaNbO_3$ from first-principles.

Perovskite oxides were the subject of numerous first-principles studies based on Density Functional Theory (DFT), most of which were performed within the usual Local Density Approximation (LDA) and, to a lesser extent, the Generalized Gradient Approximation (GGA) [11]. Although many successes have been achieved, the typical inaccuracies of these approximations impose considerable limitations. The LDA is well known to underestimate the lattice constants. While the typical underestimation of only 1 or 2 % of the experimental value is in many contexts regarded as acceptable, it reveals itself quite problematic in the study of FE oxides whose structural instabilities are extremely sensitive to volume changes. In most cases, the GGA constitutes a significant improvement over the LDA, although it has a tendency to overcorrect the LDA error for the equilibrium volume, thus leading to overestimations. In the case of ferroelectrics, for instance, the well known GGA functional of Perdew, Burke and Ernzerhof (GGA-PBE) [12] performs significantly worse than the LDA [13]. For a further description of the sensitivity of lattice distortions to volume in perovskite oxides, we refer the reader to a recent topical review [11]. It was only very recently that Wu and Cohen [14] proposed a modified GGA functional (GGA-WC) that is accurate for $BaTiO_3$ and $PbTiO_3$ and thus allows a reliable description of the structural properties of FE oxides. The complex structural behavior of $NaNbO_3$ makes this perovskite suitable for testing the limitations of the different functionals.

In this paper we report a first-principles study of $NaNbO_3$ investigating the relative stability of different phases using the LDA and GGA-WC functionals. The full phonon-dispersion curves of the cubic structure are calculated form first-principles in order to search for the presence of lattice instabilities in the whole Brilloiun zone. Finally, we report comparative calculations for the structural instabilities in $NaNbO_3$ and $KNbO_3$ at different volumes, in order to provide insights for a qualitative understanding of KNN solid solutions.

## 2. Method

The calculations were performed within DFT using the full-potential linearized augmented plane-wave ((L)APW) method with the addition of local-orbital basis functions as implemented in the WIEN2K code [15]. The LAPW and augmented plane wave with local orbitals (APW + lo) methods were adopted as basis because the highest efficiency is found for a mixed basis set. The 'physically important' l quantum numbers are treated by APW+lo [16] but the higher values of l were treated by the LAPW method [17]. Also for the low-lying valence states, semi-core states, the local orbital method was utilized. The exchange-correlation potentials used were the Perdew and Wang version of the LDA [18] and the Wu and Cohen version of the GGA [14]. The muffin-tin sphere radii $R_{MT}$ = 2.0, 1.85, and 1.5 a.u. were used for Na (K), Nb and O, respectively. All calculations were converged with respect to all parameters used, to the precision necessary to support our calculations (converged forces and total-energy differences). The value of the parameter $RK_{max}$, which controls the size of the basis set for the wave functions, was chosen to be 8, which resulted in a well-converged basis sets. Integrations in reciprocal space were performed using the tetrahedron method with 216 k points in the first Brillouin zone. Convergence tests indicate that only small changes result from going to a denser k mesh or to a larger value of $RK_{max}$.

Relaxations of trial structures for the more relevant crystallographic phases of $NaNbO_3$ (cubic with 5 atoms/unit cell, $T_2$ with 10 atoms/unit cell , $T_1$ with 40 atoms/unit cell, P with 40 atoms/unit cell and N with 10 atoms/unit cell ) were performed at different volumes. At each volume, the total energy was computed on a grid sampling the lattice parameters with a step size of 0.05 a.u., centered on the experimental lattice parameters. For each of these points, a full relaxation of the internal degrees of freedom was performed. The optimal structure determined at each volume therefore accounts for both the equilibrium lattice parameters and the equilibrium atomic coordinates inside the unit cell.

To search for the presence of structural instabilities in the full Brillouin zone, we determined the phonon dispersion relations for the cubic structure along high symmetry lines. The phonon spectrum of the system was computed by performing force calculations for selected distortions determined by the help of the Phonopy package [19], which is based on a supercell approach [20]. The LO-TO splitting, which is a general feature of polar materials, is calculated with Phonopy by a non-analytical-term correction.

Phonon frequencies and eigenvectors of $NaNbO_3$ and $KNbO_3$ (at the Γ, M, and R points) as a function of volume were calculated using the frozen-phonon method, which involves computing the forces induced by finite displacement of individual atoms via the Hellmann–Feynman theorem. From the force as a function of displacement, the dynamical matrix was constructed and diagonalized. We note that the frozen-phonon method provides only the TO ferroelectric modes. Once the eigenvectors have been determined, we evaluate the total energy as a function of the displacement pattern of the unstable (imaginary frequency) modes. This curve will provide the instability energy associated with a particular phonon mode.

## 3. Results and discussion

We present first DFT results for cubic $NaNbO_3$. While LDA gives an equilibrium lattice parameter $a_o$ = 3.9184 Å and a bulk modulus B = 206.45 GPa, GGA-WC predicts $a_o$ = 3.9516 Å and B = 193.02 GPa. The experimental value for the cubic lattice constant is $a_o$ = 3.945 Å [21]. These results are compatible with the tendency of the LDA to overbind, resulting in a short lattice constant, while GGA to expand the lattice constant [22]. We must point out, however, that the experimental value reported above was measured at 950K (the cubic phase of $NaNbO_3$ is stable above 913K), and it is not correct to compare that value with zero-temperature ab-initio calculations. It is also incorrect to compare the theoretical cubic structure with the low-temperature experimental data directly, because the structure of $NaNbO_3$ in the ground state is strongly distorted by the tilting of the oxygen octahedra. So, the validity of LDA and GGA-WC functionals must be tested by comparing the predicted theoretical structure of the ferroelectric N phase with low-temperature experimental results. Table 1 shows the theoretical results for the full-optimization of the crystal structure of $NaNbO_3$ in the ferroelectric N phase, showing that GGA-WC agrees very well with the experimental result.

Total energy versus volume data for the $T_2$ , $T_1$ , P and N phases of $NaNbO_3$ are shown in Fig. 1 for both LDA (a) and GGA-WC (b) functionals. Volume and energy are given per single $NaNbO_3$ formula unit in all cases (there are two formula units per unit cell for the $T_2$ and N phases, and eight per primitive unit cell for the $T_1$ and P structures). The energy minimum of cubic structure is taken as zero of energy in both plots. The two functionals give a relative stability of the cubic, $T_2$ and $T_1$ phases in good agreement with experiments. There are, however, differences in the energetics of the low-energy phases. The LDA functional predicts that the antiferroelectric P phase is the ground-state structure of $NaNbO_3$ with an equilibrium volume of 57.08 Å$^3$ and an energy gain of ~ 0.2 mRy/cell with respect to the ferroelectric N phase. This wrong prediction is fixed by the GGA-WC functional, where the total energies corresponding to the N and P phases, evaluated at their equilibrium volumes, are practically identical. The small energy difference between the two phases could explain the coexisting ferroelectric and antiferroelectric structures observed in experiments [1]. We thus conclude that while LDA does not reproduce properly the structural behavior of $NaNbO_3$, GGA-WC gives important improvements. We wish to point out however that since the ferroelectric phase occurs at very low temperature, it is likely that quantum fluctuations would need to be included to determine the actual low-temperature structure [24]. For example, Raevskaya et. al. [25] found that, in some single crystals, a novel quantum paraelectric state coexists with a metastable ferroelectric state at low temperatures. In the rest the paper we will use the GGA-WC functional to calculate lattice dynamics properties of the material.

The first-principles calculations of the full phonon-dispersion relations for the unstable cubic perovskite structure led to significant new insights into the lattice instabilities and local distortions in ferroelectric oxides [26]. The instabilities are manifested as phonons with negative $\omega^2$; and these imaginary frequencies are by convention plotted as negative. We then evaluate the phonon dispersion relations of cubic NaNbO3 using the so-called supercell method, as was explained in the above section. The phonon branches were calculated at the GGA-WC equilibrium volume of the cubic structure employing a 135-atom supercell. Effective atomic force-constants were obtained by computing the forces for three selected configurations. In each configuration a single atom was displaced by +0.02 Å. Na and Nb were displaced along the (1,0,0) direction, while O along (1,1,0). The supercell method employed here necessarily neglects atomic force constants between atoms at distances not contained in the supercell. This may cause, in general, inaccuracies in the calculated phonon branches, which therefore should be considered with caution. However, for the supercell we used, the method is confident at the points Γ, X, M, and R in the sense that all relevant force constants are included. The resulting phonon branches are depicted schematically in Fig. 2. The phonon branches clearly show the instability of the cubic configuration. It is instructive to compare the unstable-phonon branches with those for the classical perovskite $KNbO_3$, which are plotted in doted lines. We note first that the phonon branches obtained for $KNbO_3$ are in good agreement with those obtained by Yu and Krakauer using the linear-response approach [27]. Both $NaNbO_3$ and $KNbO_3$ present an unstable ferroelectric mode at Γ, characterized by the displacement of the Nb ion against the oxygen octahedron. An analysis of the eigendisplacements corresponding to the $\Gamma_{15}$ phonons (Table 2) shows that the Na atoms are indeed involved in the motion in such a way that both atoms Na and Nb move out of phase with respect to the oxygen atoms; so the Na atom contributes significantly to the ferroelectric $\Gamma_{15}$ mode in $NaNbO_3$. These results are similar to the behavior found for $ATiO_3$ perovskites in the sense that as the A atoms decrease in size, their contribution to the ferroelectric motion increases and becomes dominant, whereas the antiphase motion involves not O|| but O⊥ displacements [28].

The ferroelectric instability is not restricted to the Γ point. In both perovskites branches of polar unstable modes extend over much of the Brillouin zone, and one of the two unstable modes at Γ remains unstable at the Γ-X-M planes. If we move away from this plane towards the R point, there is a rapid stiffening of this mode. These features confine the $\Gamma_{15}$ instability to three quasi-two-dimensional "slabs", which are perpendicular to each other and they intersect at the Γ point. This behavior in the reciprocal space determines the existence of an unstable finite chain of Nb displacements in real space, similarly to what is observed in $KNbO_3$ [27]. Except for these modes, all the other phonons are stable in $KNbO_3$ , which makes the behavior of the unstable branches relatively easy to understand [27]. In $NaNbO_3$, however, there are additional unstable branches. Beside the ferroelectric ($\Gamma_{15}$) instabilities, there are rotational instabilities ($R_{25}$, $M_3$ and the branch connecting them) which are characterized by oxygen octahedral tilts. In the $R_{25}$ mode, the oxygen octahedra rotate about one of the cubic axis containing the Nb atom and the sense of rotation is opposite in adjacent cells. The $M_3$ mode is nearly identical to the $R_{25}$ mode but the rotation of the octahedra is in the same sense in neighboring cells along the z axis.

The regions of instability in the full Brillouin zone are depicted in Figure 3, where reciprocal space regions with at least one imaginary frequency mode are colorated in grey. The three slab-like portions of $\Gamma_{15}$ instabilities can be clearly visualized in the figure. Regarding oxygen octahedra rotations, it is clear from the dispersion relations that unstable tilting modes are present along the entire R-M-R edge, generating (in a repeated-zone scheme) an instability region consisting of eight cylindrical tubes with axes along the Brillouin zone edges. The essentially 1-D character of the cylinder-like regions corresponds to direct-space planar instabilities characterized by rotations of oxygen octahedra, as occurs in $SrTiO_3$ [29, 30]. We note that our theoretical results give support to x-ray diffuse scattering experiments in $NaNbO_3$, where two types of distinct diffuse scattering were observed in {100} reciprocal planes and on (100) reciprocal axes [31] . While the former can be attributed to chains of Nb atoms oriented along the (100) direction, the second one is generated from planar disorder attributed to rotations of oxygen octahedra. Both types of disorder exist in the cubic paraelectric phase.

The dispersion curves showed in Figure 2 indicate that the mode with larger imaginary frequency is at the center of the Brillouin zone, that is $|\omega(\Gamma_{15})| > |\omega(R_{25})| = |\omega(M_3)|$. This inequality is reversed for the corresponding energy gains, which were evaluated from total energy calculations as a function of the corresponding eigendisplacement patterns at different mode amplitudes. The results are plotted in Figure 4, where the energy per formula unit is plotted as a function of a magnitude characteristic of each mode. While the $\Gamma_{15}$ mode displays a ferroelectric instability with energy gain of 1.2 mRy/ cell (Fig.4-a), the energy gains of the $R_{25}$, and $M_3$ modes are considerable larger, 2.6 and 2.4 mRy/ cell, respectively (Fig.4-b). This explains why octahedral tilts corresponding to the $R_{25}$ and $M_3$ modes can be considered as primary order parameters, inducing the first two structural transitions, from the cubic into the $T_2$ and $T_1$ phases.

In perovskite oxides the rotational and ferroelectric instabilities have an opposite dependence on volume: increase in volume tends to destabilize the polar mode, while stiffening the rotational mode [11]. Thus, this competition is quite sensitive to changes in the lattice constant by external stress or chemical substitution, which allows some anticipation of the solid-solution phase diagrams on the basis of the endpoint compounds [30]. So, how the competition between the $\Gamma_{15}$ , $R_{25}$ and $M_3$ modes of $NaNbO_3$ and $KNbO_3$ behaves as a function of volume is a relevant information for the qualitative understanding of the structural behavior of $K_{1-x}Na_xNbO_3$ solid solutions. To this end, we calculated the phonon frequencies of both materials at different volumes using the frozen phonon approach. The results are plotted in Figure 5, where the vertical lines indicate the equilibrium lattice constants for cubic $NaNbO_3$ ($a_o$ = 3.9516 Å) and $KNbO_3$ ( $a_o$ = 3.9916 Å ), delimiting the range of volumes of interest to the $K_{1-x}Na_xNbO_3$ solid solutions. As shown in Figure 5, the $\Gamma_{15}$ phonons keep the same variation trend in both materials. The square of such frequencies are always negative, which indicates that both oxides have a ferroelectric instability and thus cannot be paraelectric for any pressure at 0K (unless quantum fluctuations stabilize the cubic phase) . The magnitude of this frequency square first increases with decreasing volume and then decreases below a critical lattice constant, confirming that these systems have to become more ferroelectrics above a critical pressure. We note that the aforementioned trend is not only restricted to the present compounds but also appears in other perovskites [32]. On the other hand, the rotational modes of the two materials behave differently for volumes in the range of interest to the KNN solid solutions. While the $M_3$ and $R_{25}$ modes are unstable for $NaNbO_3$ , they have positive frequencies for $KNbO_3$ at the same range of volumes (in this material the antiferrodistortive mode becomes unstable for lattice constants less than ~3.87 Å). This result indicates that the cell volume contraction is not a relevant parameter for the destabilization of rotational modes in $K_{1-x}Na_xNbO_3$ solid solutions.

For that reason, we also investigated the effect of different Na and K arrangements on the stability of the $M_3$ rotational mode in $K_{1-x}Na_xO_3$. The calculations were made at a fixed volume, which was chosen as the equilibrium volume of $NaNbO_3$. The stability of the rotational mode is expressed in terms of the stiffness $\kappa = \frac{1}{2}\frac{\partial^2 E}{\partial \Phi^2}$, where $\Phi$ is the rotation angle of the oxygen octahedra. This magnitude was calculated for different cation arrangements on 2x2x2 supercells with 40 atoms/cell. For each concentration, all the possible configurations over the eight available A-sites were considered. From these data, the average stiffness is computed. Figure 6 displays the resulting stiffness as function the Na concentration x. It is clear that the addition of Na on the $KNbO_3$ lattice decreases monotonically the stiffness of the rotational mode, which becomes negative for $x > 0.5$. This means that the rotational mode in $K_{1-x}Na_xNbO_3$ solid solutions is unstable for $x > 0.5$, which is in agreement with an updated phase diagram reported recently [10], where the low-temperature *R3m* structure of $KNbO_3$ transforms into the tilted *R3c* phase at x ~0.5.

In summary, we found that LDA underestimates the equilibrium volume of $NaNbO_3$ and predicts an incorrect ground state for this oxide. The GGA-WC functional, on the other hand, significantly improves the equilibrium volume and provides relative phase stability in better agreement with experiments. The ab-initio calculation of phonon dispersion curves allowed the determination of regions of polar and rotational instability in the Brillouin zone, which support x-ray diffuse scattering experiments [32]. Comparative calculations of phonon instabilities in $NaNbO_3$ and $KNbO_3$ provided insights for the understanding of the structural behavior of $K_{1-x}Na_xNbO_3$ solid solutions, demonstrating that the cell volume is not a relevant parameter for the destabilization of rotational modes. Finally, we note that the results reported here can be used for the development of atomistic models [33] to determine the temperature-driven phase transition sequence in NaNbO3 and $K_{1-x}Na_xNbO_3$ from a first-principles-based approach, independently of experimental data.

*Acknowledgements:* This material is based upon work supported by Consejo Nacional de Investigaciones Científicas y Tecnológicas de la República Argentina (CONICET) and Agencia Nacional de Investigación Científica y Tecnológica (FONCyT). M.G.S. thanks support from Consejo de Investigaciones de la Universidad Nacional de Rosario (CIUNR).

**FIGURE CAPTIONS**

**Figure 1:** Energy as a function of volume for the $T_2$, $T_1$, P and N phases of $NaNbO_3$ from ab-initio calculations using (a) LDA and (b) GGA-WC functionals. The zero energy corresponds to the energy minimum of the cubic phase. Volume and energy are per single $NaNbO_3$ formula unit in all cases (there are two formula units per unit cell for the $T_2$ and N phases, and eight per primitive unit cell for the $T_1$ and P structures).

**Figure 2:** Calculated phonon dispersion curves for cubic $NaNbO_3$ at the theoretical lattice constant. For comparison, the imaginary frequency modes for cubic $KNbO_3$ are plotted in dashed lines.

**Figure 3:** Phonon instability regions (colorated in grey) over the entire Brillouin zone, which presents at least one imaginary frequency mode. The $\Gamma_{15}$ instability region consists of three quasi-two-dimensional slabs, which are perpendicular to each other and intersect at Γ. The region with rotational instabilities is formed by eight narrow cylindrical tubes with axes along the Brillouin zone edges.

**Figure 4**: Total energy as a function of the $\Gamma_{15}$ ferroelectric mode displacement (a), and the oxygen octahedra rotation angle (b). The calculations were made at the equilibrium lattice constant of cubic $NaNbO_3$ obtained by using the GGA-WC functional.

**Figure 5**: Frequencies of the $\Gamma_{15}$, $R_{25}$, and $M_3$ modes as a function of lattice constant in (a) $NaNbO_3$ and (b) $KNbO_3$. The vertical lines indicate the equilibrium lattice constants for cubic $NaNbO_3$ ($a_o$ = 3.9516 Å) and $KNbO_3$ ( $a_o$ = 3.9916 Å ), delimiting the range of interest to the $K_{1-x}Na_xNbO_3$ solid solutions.

**Figure 6**: Stiffness of the $M_3$ rotational as function x in $K_{1-x}Na_xNbO_3$ solid solutions, indicating that this rotational mode is unstable for $x > 0.5$.

Table 1: Lattice parameters and internal coordinates for $NaNbO_3$ in the ferroelectric N phase (R3c) obtained with LDA (first row) and GGA-WC (second row) functionals, and compared with experiments (third and fourth rows).

| Lattice parameters | | |
|---|---|---|
| a(Å) | | 5.496<br>5.570<br>5.566 [a]<br>5.552 [b] |
| α(deg) | | 59.15<br>59.15<br>59.05 [a]<br>59.16 [b] |
| Atom | Wyckoff pos. | x y z |
| Na | 2a | 0.271 0.271 0.271<br>0.272 0.272 0.272<br>0.272 0.272 0.272 [a]<br>0.272 0.272 0.272 [b] |
| Nb | 2a | 0.010 0.010 0.010<br>0.016 0.016 0.016<br>0.016 0.016 0.016 [a]<br>0.016 0.016 0.016 [b] |
| O | 6b | 0.322 0.177 0.742<br>0.318 0.191 0.749<br>0.357 0.143 0.750 [a]<br>0.320 0.183 0.747 [b] |

a- Reference [23]. Measured at 123K.

b- Reference [1]. Measured at 12K.

Table 2: Eigendisplacement of the $\Gamma_{15}$ ferroelectric mode in $NaNbO_3$ and $KNbO_3$ using the GGA-WC functional. The calculations were made at the equilibrium lattice constant of the cubic structures. The displacement vector is obtained from the eigenvector by dividing each component by the square root of the corresponding atomic mass and then normalizing to unity.

| Atom | $NaNbO_3$ | $KNbO_3$ |
|---|---|---|
| Na/K | -0. 258 | 0. 029 |
| Nb | -0. 216 | -0. 296 |
| O‖ | 0. 468 | 0.  624 |
| O ⊥ | 0. 578 | 0. 511 |

*Figure-1, Machado et al, Physical Review B*

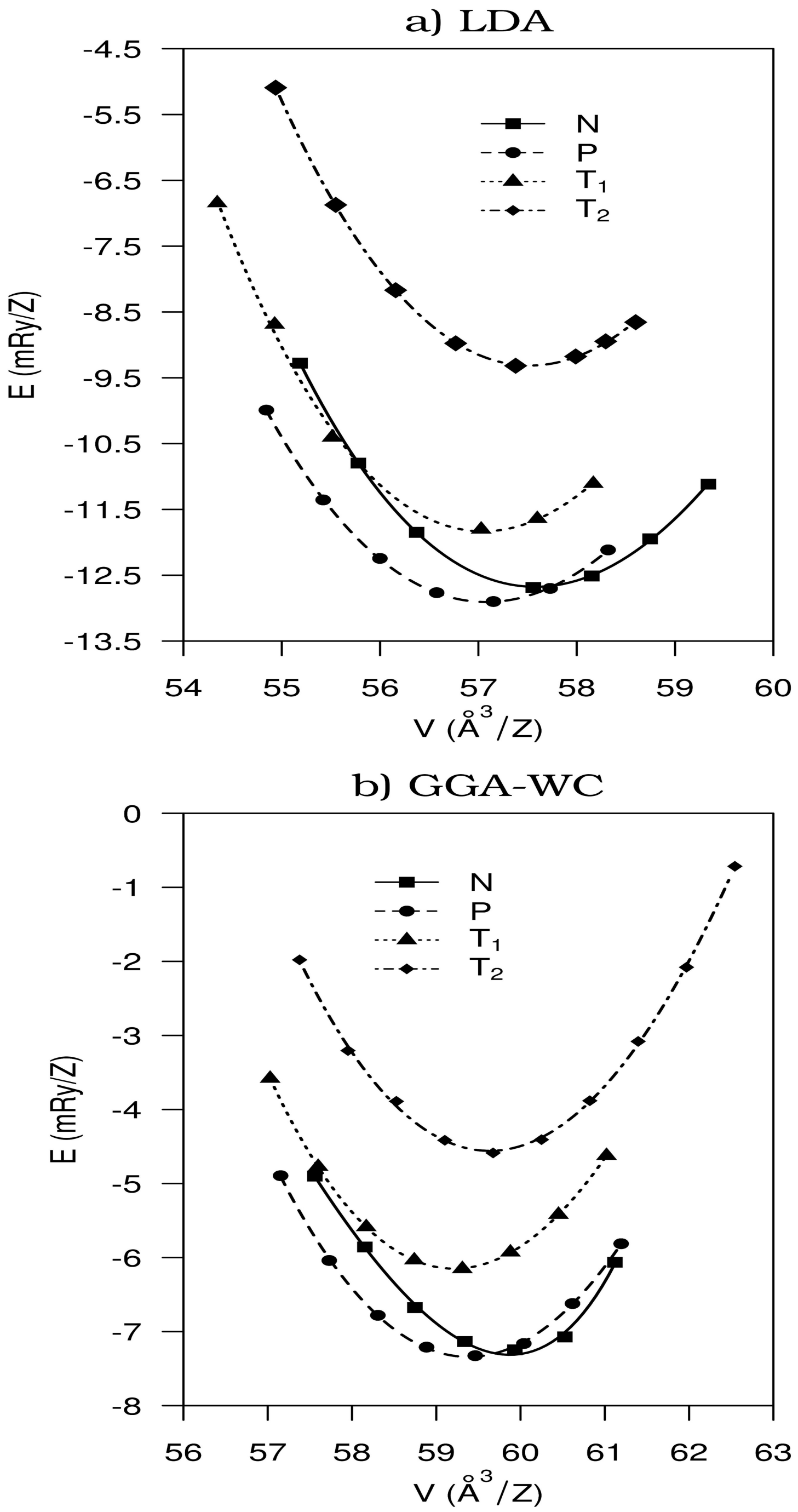

*Figure-2, Machado et al, Physical Review B*

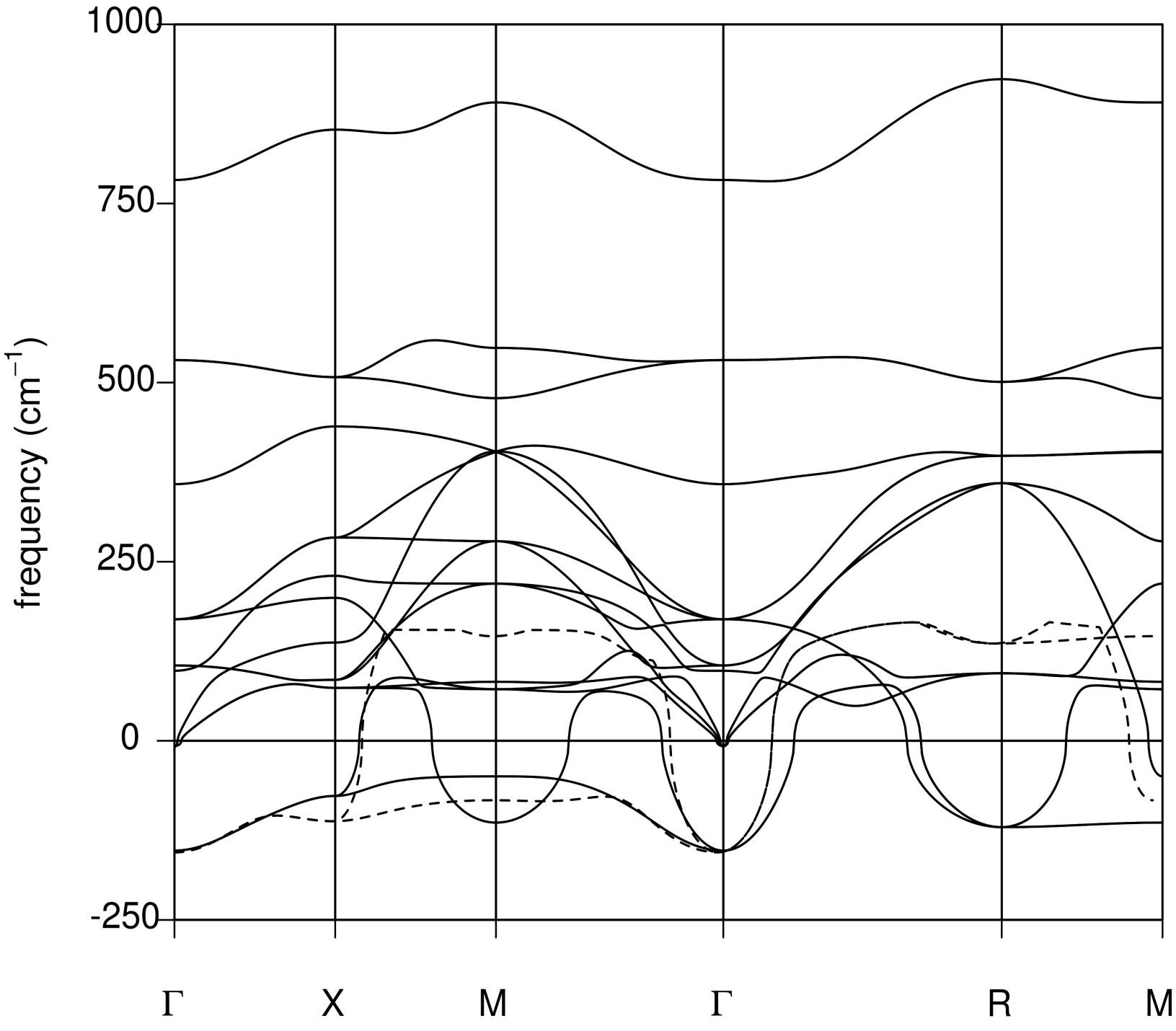

Figure-3, Machado et al, Physical Review B

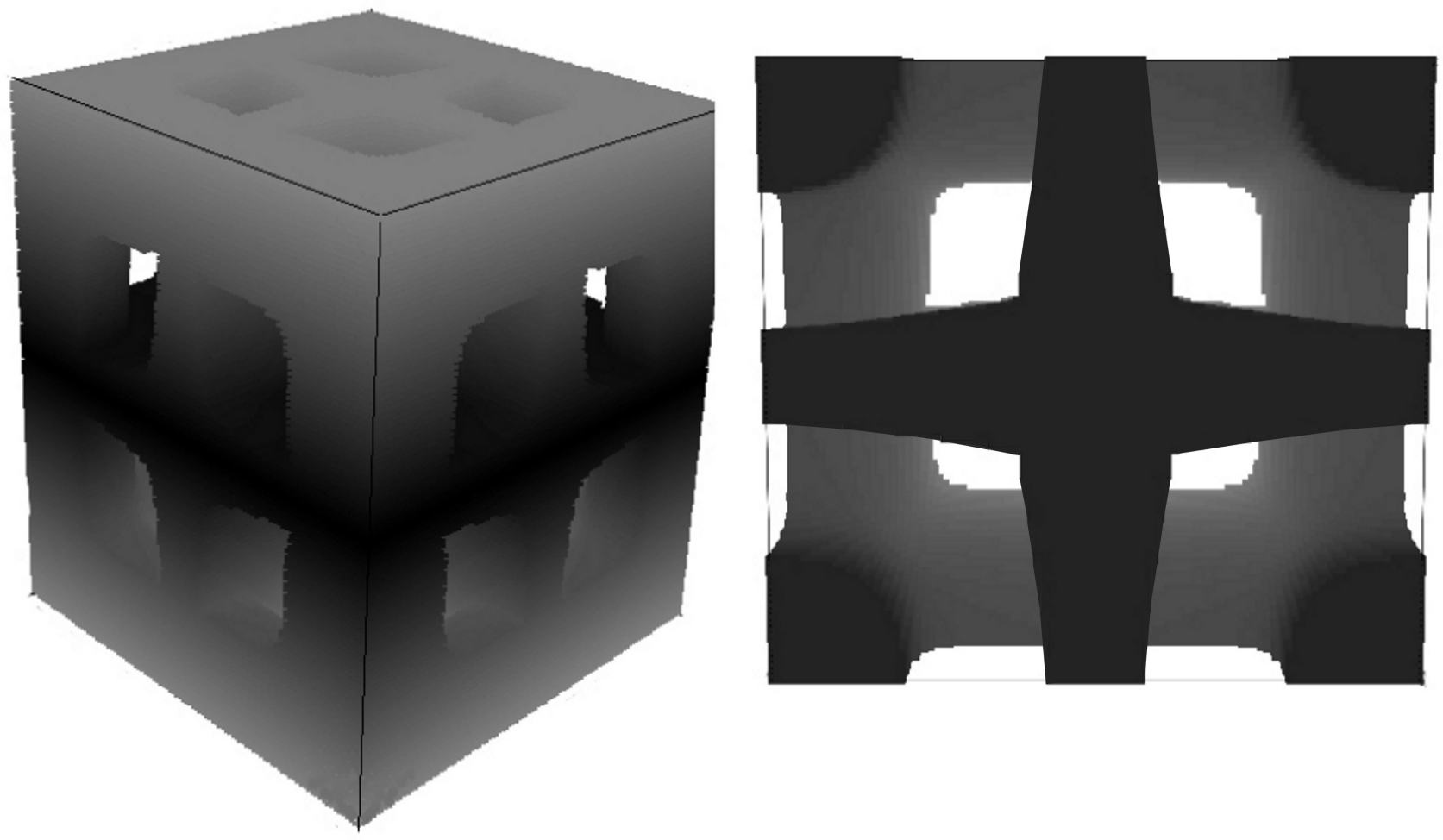

*Figure-4, Machado et al, Physical Review B*

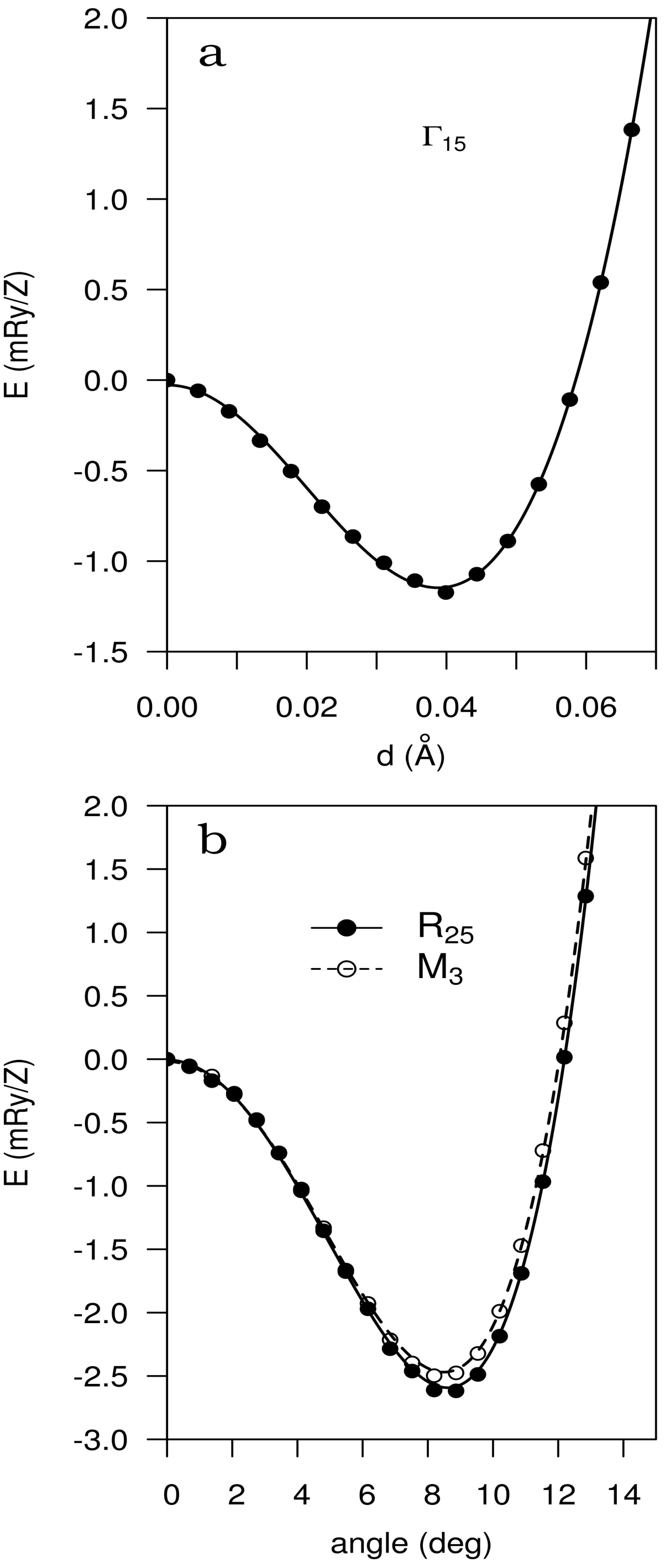

*Figure-5, Machado et al, Physical Review B*

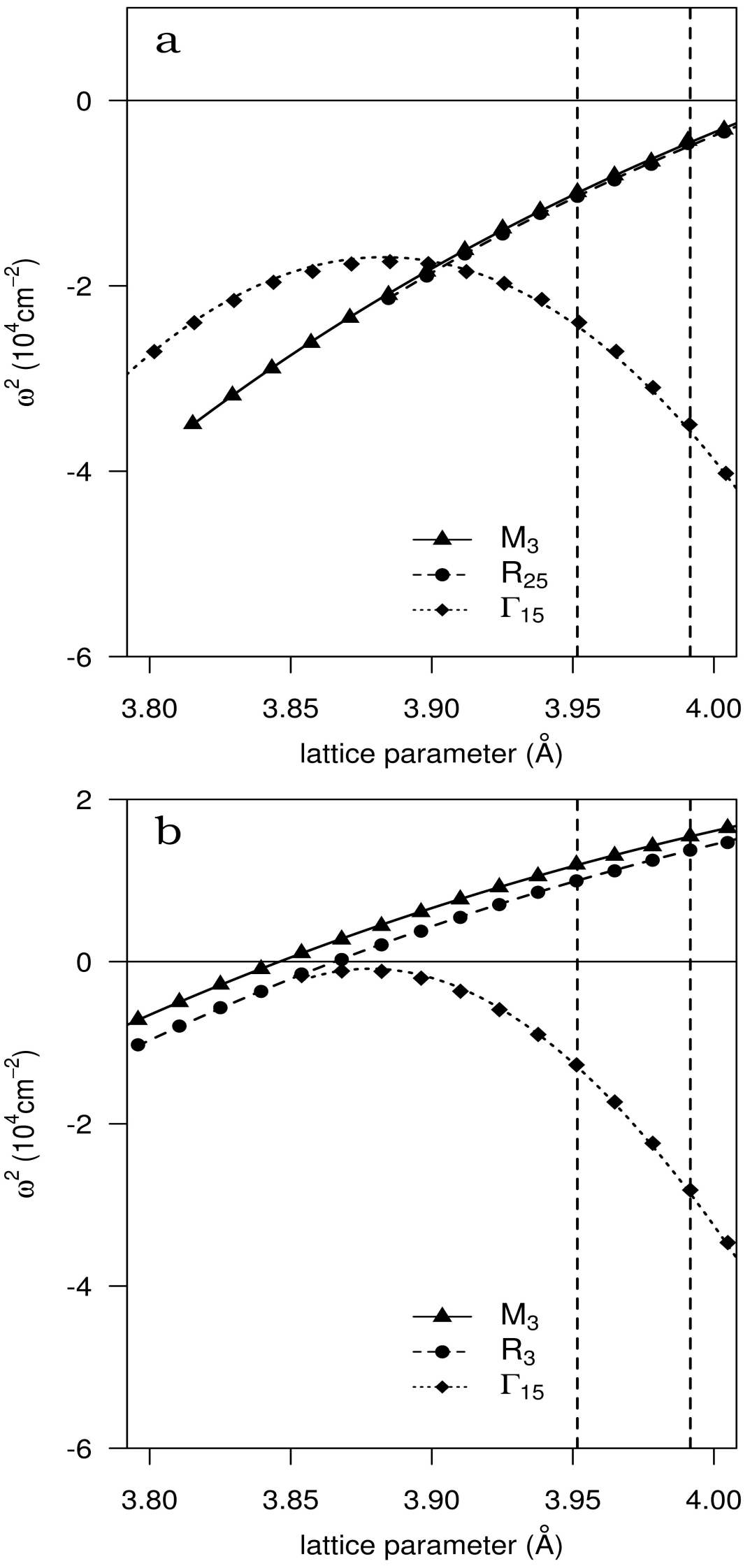

*Figure-6, Machado et al, Physical Review B*

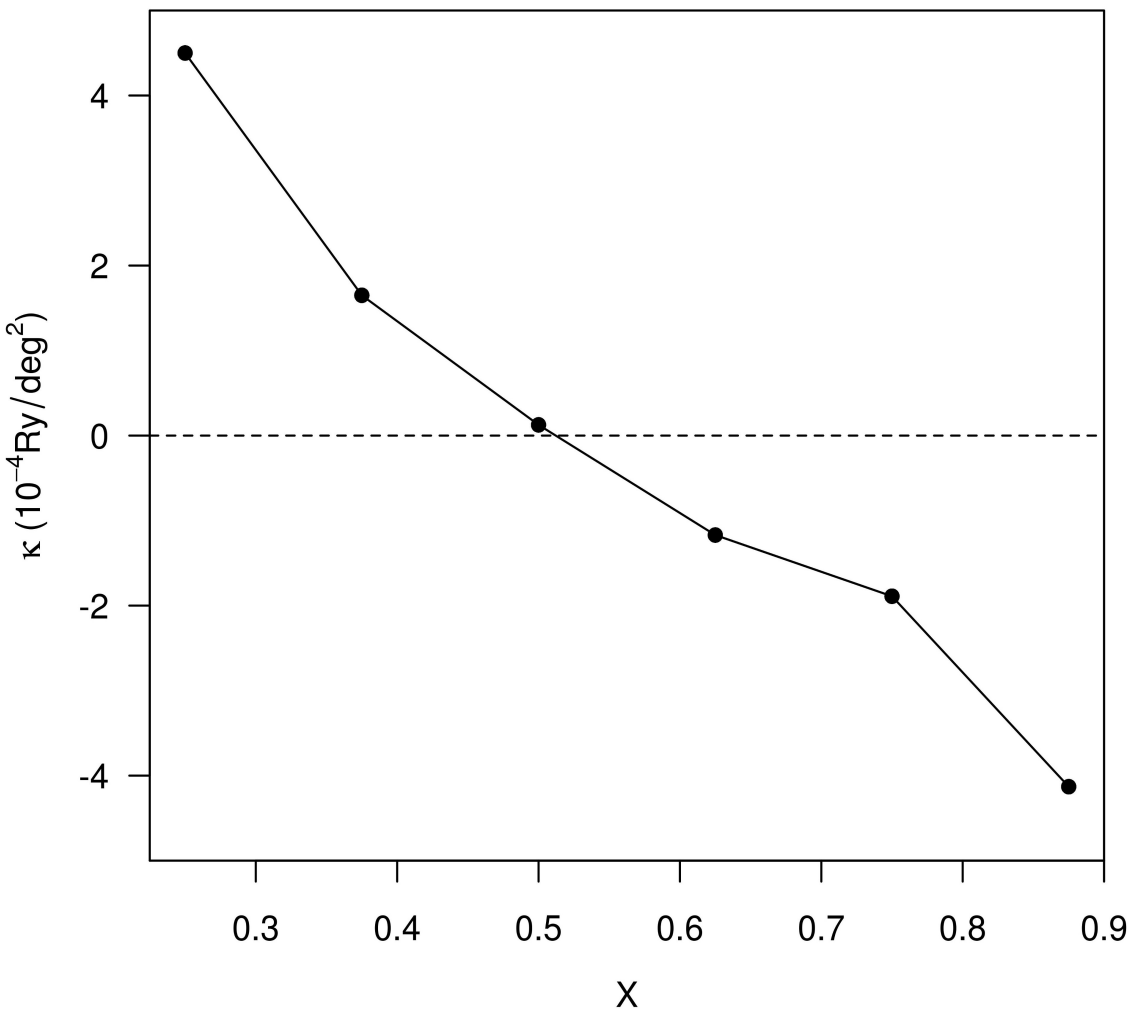